\documentstyle[aps,preprint]{revtex}

\begin{document}

\title{Widths of Isobaric Analog Resonances: a microscopic approach}

\author{G. Col\`o$^1$, H. Sagawa$^2$, N. Van Giai$^3$, P.F. Bortignon$^1$, 
	and T. Suzuki$^4$}

\address{$^1$ Dipartimento di Fisica, Universit\`a degli Studi and INFN, 
	 Sezione di Milano, Via Celoria 16, 20133 Milano, Italy}

\address{$^2$ Center for Mathematical Sciences, The University of Aizu, 
	 Aizu-Wakamatsu, Fukushima 965, Japan}

\address{$^3$ Division de Physique Th\'eorique, Institut de Physique
	 Nucl\'eaire, 91406 Orsay Cedex, France}

\address{$^4$ Department of Physics, College of Humanities and Sciences, 
	 Nihon University, Sakurajosui 3-25-40, Setagaya-ku, 
	 Tokyo 156, Japan}

\date{\today}

\maketitle

\smallskip

\begin{abstract}
A self-consistent particle-phonon coupling model is used to 
investigate the properties of the isobaric analog resonance in $^{208}$Bi. 
It is shown that quantitative 
agreement with experimental data for the energy and the width 
can be obtained if 
the effects of isospin-breaking nuclear forces are included, in 
addition to the Coulomb force effects. A connection between microscopic 
model predictions and doorway state approaches 
which make use of the isovector monopole resonance, is established 
via a phenomenological ansatz for the optical potential.
\end{abstract}

\newpage

\section{Introduction}
Since its discovery some 35 years ago, 
the isobaric analog resonance (IAR) has always attracted 
a considerable interest because it is one of the 
clearest manifestations of approximate isospin symmetry in atomic 
nuclei. The main causes of isospin symmetry breaking are essentially 
the long-range Coulomb force, and to a lesser extent the short-range 
nuclear forces of the charge-symmetry breaking (CSB) type and 
charge-independence breaking (CIB) type~\cite{hen}. 
Quoting a sentence from a review paper on 
this subject~\cite{Aue83}, the Coulomb interaction ``is strong enough so 
that its effects are quite easily detectable and weak enough not to destroy
or alter considerably what the nuclear force has produced.'' 
Despite this favourable situation,
we are still missing a microscopic model which is able to reproduce one  
of the physical observables related to the breaking of isospin symmetry in
nuclei, namely the total width of the IAR, 
taking into account all the isospin-breaking effects. \par
The general features of the IAR are qualitatively well understood. In 
the absence of isospin-breaking effects, the analog state would be 
degenerate with the ground state of the ($N,Z$) parent nucleus and it 
would have zero width. The effect of the Coulomb and other 
isospin-breaking forces is to shift up the analog state by several MeV 
(the so-called Coulomb displacement energy). However, these 
isospin-breaking forces do not induce strong isospin mixing of the 
states. The analog state, just as its parent state, has practically 
pure $T=T_0$ isospin ($T_0 \equiv (N-Z)/2$) whereas the neighbouring 
states in the daughter nucleus have mostly $T=T_0 - 1$. Thus, a simple 
argument based on the Fermi golden rule shows that 
a small 
spreading width for the IAR must be expected. In 
addition, the escape width must also be small because the main escape 
channels are those of low energy protons. \par 
When one attempts to build a microscopic model of IAR, the requirement 
that isospin symmetry must be restored if isospin-breaking forces are 
switched off should be taken into account. It is known that in $N \ne Z$ 
systems the Hartree-Fock (HF) approximation introduces a spurious 
isospin symmetry breaking and that a self-consistent charge-exchange random 
phase approximation (RPA) restores this broken symmetry~\cite{Lan_find}. 
By self-consistent RPA we mean that the HF single-particle spectrum 
and the residual particle-hole interaction are derived from the same 
effective two-body force. Therefore, we shall consider here only this 
self-consistent framework. 
Charge-exchange RPA was applied for the first 
time to IAR studies in Ref.~\cite{Aue80} (see also~\cite{Aue83} and 
references therein). Calculations of other types of excitations using 
charge-exchange RPA can be found in Ref.~\cite{Aue83b}.
However, the 
RPA description can at best give 
information on escape widths if continuum effects are included but it 
cannot shed any light on spreading widths because this spreading is 
due to states beyond the one particle-one hole (1p-1h) space. An 
extension of the model space to include 2p-2h configurations and 
leading to second RPA would be a more appropriate scheme. In heavy 
nuclei where the neutron excess is large one can safely replace the 
RPA by the Tamm-Dancoff approximation (TDA). In Ref.~\cite{Ada_find} 
such a second TDA calculation was performed for the IAR in $^{208}$Bi 
and it gave a reasonable estimate of the escape width 
$\Gamma^{\uparrow}$ and spreading width $\Gamma^{\downarrow}$. More 
recently, calculations based on a particle-phonon coupling 
model~\cite{Col94} also 
led to satisfactory values of $\Gamma^{\uparrow}$ and $\Gamma^{\downarrow}$ 
but some discrepancies with experimental data still remain. In the above 
mentioned works, only isospin breaking due to the inclusion of the Coulomb 
force in the HF mean field was 
considered. \par
The purpose of the present paper is twofold. Firstly, we wish to 
examine the effects of the isospin-breaking nuclear forces on the 
properties of the IAR. It is known that the CSB and 
CIB interactions affect the mass number
dependence of the Coulomb displacement energy anomalies and bring them 
in better agreement with experiment~\cite{Sag_1}. Moreover, these 
isospin-breaking nuclear forces lead to an increase of about 50\% of 
the isospin mixing in nuclear ground states~\cite{Sag_2}. 
Thus, one should expect also some sizable effects on the values of 
$\Gamma^{\downarrow}_{IAR}$. We study these effects within the particle-phonon 
coupling model~\cite{Col94} together with the short range 
parametrizations of the CSB and CIB forces of Ref.~\cite{Sag_2}. 
The calculations are performed for the typical case of $^{208}$Bi. 
Secondly, it is instructive to study the connection between a 
microscopic model such as the present one or the second TDA model of 
Ref.~\cite{Ada_find}, and IVMR doorway state approaches~\cite{Mek70,Suz96} 
where the shift and width of the IAR result from the coupling 
via the Coulomb interaction of an 
ideal analog state with the isovector monopole resonance (IVMR) 
playing the role of a doorway state. 
Here, we show that this connection can be 
established if one makes a phenomenological ansatz for the isospin 
dependence of the nucleon optical potential. \par
In Sect. II the microscopic model is presented and its results are 
discussed in Sect. III. The connection between the microscopic model and 
the approach of Ref.~\cite{Suz96} is shown in Sect. 
IV. Conclusions are drawn in Sect. V.

\section{The microscopic model}
The RPA extended so as to include the continuum coupling and 
the particle-phonon coupling has 
been described in detail in Ref.~\cite{Col94} and therefore, we shall 
first recall only the main features of the model and then proceed to 
the specific points of the present calculation.

Starting from an effective Hamiltonian $H$ with two-body Skyrme 
interaction~\cite{Bei75,SGII}, Coulomb interaction and CSB-CIB 
interactions which will be described below, 
the HF equations determine the self-consistent mean field of the 
parent nucleus. This mean field is diagonalized on a 
basis of 15 shells of harmonic oscillator wave functions ($\hbar
\omega_{osc}$ = 6.2 MeV for $^{208}$Pb). 
This procedure provides a discrete set of levels (occupied and 
unoccupied).  We select all occupied levels, and 6 unoccupied levels 
for each value of $(l,j)$ with increasing values of the radial 
quantum number $n$. This determines the subspace $Q_1$ of discrete 
1p-1h (proton particle-neutron hole) configurations. 

To account for the escape width $\Gamma^\uparrow$ 
and spreading width $\Gamma^\downarrow$ of the IAR, we build two
other subspaces $P$ and $Q_2$. The space $P$ is made of particle-hole 
configurations where the particle is in an unbound state orthogonal to 
all the above discrete single-particle levels. 
The method to calculate these unbound states is described in 
Ref.\cite{Col94}. On the other 
hand, the space $Q_2$ is built with the main configurations which are 
known to play a major role in the damping process of nuclear giant 
resonances: these configurations are 1p-1h
states coupled to a collective vibration. We have included 
in our model space all the 
isoscalar vibrations of multipolarity 2$^+$, 3$^-$ and 4$^+$ up to the
energy of 20 MeV and which exhaust more than 1\% of the 
energy-weighted sum rule (EWSR) of the corresponding multipole 
operator. 
These collective vibrations are calculated consistently in HF-RPA 
within the $Q_1$ space and 
they constitute the phonons of our particle-phonon coupling model.  

Using the projection operator formalism one can easily find that the 
effects of coupling the subspaces $P$ and $Q_2$ to $Q_1$ are described 
by the following effective Hamiltonian acting in the $Q_1$ space: 
\FL
\begin{eqnarray}
     \ & {\cal H} & (E) \equiv Q_1 H Q_1 + W^\uparrow(E)
     + W^\downarrow(E) \nonumber \\
     = & Q_1 & H Q_1 + Q_1 H P {\textstyle 1 \over \textstyle
     E - PHP + i\epsilon} P H Q_1
     + Q_1 H Q_2 {\textstyle 1 \over \textstyle
     E - Q_2 H Q_2 + i\epsilon}
     Q_2 H Q_1, \nonumber \\
     \ & \ &
\label{H_eff}\end{eqnarray} 
where $E$ is the excitation energy. 
For each value of $E$ the RPA equations corresponding to 
this effective, complex Hamiltonian ${\cal H} (E)$ are solved. The resulting 
sets of eigenstates enable one to calculate all relevant 
quantities such as giant resonances energies and widths (see Ref.~\cite{Col94}). 
In practice, we use the HF-RPA 
states (corresponding to positive and negative 
eigenvalues) as a basis for the $Q_1$ space because we can truncate 
this basis and neglect states which have negligible $T_{-}$ strength. 

A simplifying approximation is made when calculating the matrix elements of 
$W^\downarrow$ by neglecting the interactions among the states within 
$Q_2$. 
Each matrix element $W^\downarrow_{ph,p^\prime h^\prime}$ is a sum of 
four terms whose 
diagrammatic representation is shown in Fig. 1. To evaluate these diagrams 
we use the following expression for the
particle-vibration vertices:  
\begin{equation}
 V = \sum_{\alpha\beta} \ \sum_{LnM} \langle \alpha |
 \varrho^{(L)}_n (r) v(r) Y_{LM}(\hat r) | \beta \rangle
 \ a^\dagger_\alpha a_\beta.
\label{pvc}\end{equation} 
This form comes from 
the particle-vibration coupling model where the vibration (phonon) $|n\rangle$ 
is characterized by its angular momentum $L$ and its radial transition density
$\varrho^{(L)}_n(r)$. 
The form factor $v(r)$ appearing in (\ref{pvc}) is related to the
particle-hole interaction derived from the Skyrme force by
$V_{ph}(\vec r_1,\vec r_2)=v(r_1)\delta(\vec r_1 -\vec r_2)$.  
When deriving
this form factor, the velocity-dependent terms of the residual Skyrme 
interaction are dropped.  

It was pointed out in Ref.\cite{Col94} that the particle-phonon 
coupling model does not automatically insure the isospin symmetry 
properties of the nuclear forces (contrarily to a fully 
microscopic 2p-2h model like in Ref.\cite{Ada_find}). One must 
therefore enforce isospin symmetry in the evaluation of 
$W^\downarrow_{ph,p^\prime h^\prime}$ by an appropriate isospin 
projection procedure\cite{Col94}. 

In the present calculations the Skyrme 
interactions SIII~\cite{Bei75} and SGII~\cite{SGII} have been employed 
for the isospin symmetric part of the Hamiltonian.  
In addition, CSB and CIB effective nucleon-nucleon forces determined 
in Ref.\cite{Sag_2} are also included. These forces were obtained from 
a short-range expansion of Yukawa-type interactions and they have a 
form similar to that of Skyrme forces:
\begin{eqnarray}
 V_{CSB} = & {1\over 4} & \{\tau_z(1) + \tau_z(2)\} \{s_0 (1+y_0 P_{\sigma})
	   + {1\over 2}s_1 (1+y_1 P_{\sigma}) \hfill\nonumber \\
    \      & \times & ({\vec k}^2 + {\vec k}^{\prime 2}) 
	   + s_2 (1+y_2 P_{\sigma}) 
	   {\vec k}^\prime \cdot {\vec k} \}, 
\label{V_CSB}\end{eqnarray}
and 
\begin{eqnarray}
 V_{CIB} = & {1\over 2} & \tau_z(1)\tau_z(2) \{u_0 (1+z_0 P_{\sigma})
	   + {1\over 2}u_1 (1+z_1 P_{\sigma}) \hfill\nonumber \\
    \      & \times & ({\vec k}^2 + {\vec k}^{\prime 2}) 
	   + u_2 (1+z_2 P_{\sigma}) 
	   {\vec k}^\prime \cdot {\vec k} \}.
\label{V_CIB}\end{eqnarray}
The parameters $s_i$ and $u_i$ are given in Ref.~\cite{Sag_2}, and 
all exchange parameters $y_i$ and $z_i$ are -1 because of the 
singlet-even character of $V_{CSB}$ and $V_{CIB}$. Therefore, they do 
not contribute as residual particle-hole interactions in the isovector 
channel and their only influence is through their contributions to the 
HF mean field.

\section{Discussion of results} 
Three types of calculations have been done which are labeled by 
I, II, III respectively. In calculation I 
the starting Hamiltonian contains only the Skyrme interaction 
without the Coulomb force between protons and without 
CSB-CIB nuclear forces. In calculation II the Coulomb interaction is 
also included. 
In calculation III the CSB-CIB forces are added. Thus, the three
calculations have an increasing degree of isospin breaking and they 
are expected 
to lead to increasing values of the IAR width 
according to the arguments recalled before. In calculations II and III 
the Coulomb exchange contributions to the mean field are treated in 
the Slater approximation whereas the Coulomb p-h residual interaction 
is dropped. The 
results obtained by using respectively the interactions 
SIII and SGII are shown in Tables 1 and 2. 
Calculation II is in principle equivalent to what was done 
in Ref.~\cite{Col94} except for two changes in the model spaces 
$P$ and $Q_2$, and in the 
averaging parameter $\Delta$. Because of these changes, the values of 
the mean energies and widths of the 
IAR we quote as calculation II are slightly different from the correspnding 
values reported in~\cite{Col94}. The spaces $P$ and $Q_2$ 
have been enlarged 
with respect to the calculation of Ref.~\cite{Col94} by including 
a larger (96 instead of 64) number of 
unoccupied proton single-particle states. 
This affects mainly the escape and spreading widths calculated with the
interaction SIII which both increase by about 10 keV whereas the widths 
calculated with the interaction SGII are less affected. Also, the 
averaging parameter $\Delta$ (see Ref.~\cite{Col94}) employed in the 
calculation has 
been changed in the present work from 100 keV, as it was in~\cite{Col94}, to 
200 keV. 
In fact, we have carefully studied the dependence of the spreading width 
on the parameter $\Delta$, and we have found that 
$\Gamma^\downarrow$ increases by about 20 keV when $\Delta$ is 
changed from 100 keV to 200 keV (as it was 
already discussed in Ref.~\cite{Col94}) and 
also increases by another 20 keV when $\Delta$ is changed 
from 200 
keV to 400 keV. On the other hand, $\Gamma^\downarrow$ remains constant if 
$\Delta$ is set above 400 keV. Therefore, we have adopted the value of 
200 keV for $\Delta$ since this value is intermediate between 0 keV 
and 400 keV (the value at which $\Gamma^\downarrow$ saturates). This study 
allows us to say that the uncertainity on the values of $\Gamma^\downarrow$ 
due to the freedom of choice of the parameter $\Delta$ is about $\pm$ 20 
keV.  

The calculation without any isospin-breaking force (calculation I) has 
been performed to show that our procedure can recover the isospin symmetry 
reasonably well at each step. 
The first
step (a)  is  discrete TDA since discrete 
RPA is not possible in this case because
of the negative energy configurations.  
Indeed, without the Coulomb interaction the neutron states lie higher 
in energy than their 
proton partners and the energies $\varepsilon_p - \varepsilon_h$ of the 
unperturbed proton particle-neutron hole 0$^+$ configurations are 
negative. For the six excess neutron hole levels and the 
corresponding proton particle levels, these 
values are between -7.3 MeV and -8.3 MeV in the case of the force SGII 
and between -8.6 MeV and -10 MeV in the case of the force SIII. 
In an ideally accurate numerical 
calculation these negative energy configurations would be coherently 
pushed up to form a collective state at zero energy. Small numerical 
inaccuracies can weaken the residual interaction, and as a result complex RPA 
eigenvalues may appear. 
This can be easily understood in the simple case of 
a schematic model, or in a case containing a single state, where the RPA and 
TDA energies are related by $E_{RPA}^2 = E_{TDA}^2 - V^2$ with $V$ 
representing the interaction term. In the limit $V_C=0$ the r.h.s. may  
become slightly negative and consequently $E_{RPA}$ is imaginary, 
if $E_{TDA}^2$ and $V^2$ do not exactly cancel 
numerically. This difficulty of complex RPA solutions in the case 
$V_C=0$ is well known and, for example, in Ref.~\cite{Aue83} the way out was 
to slightly renormalize the nuclear residual interaction. 
Therefore, only TDA  
is possible 
for the study of the IAR without the Coulomb interaction. 
One should also 
note that, contrary to a naive view, adding the Coulomb interaction does not 
result in an overall shift of the RPA eigenvalues. This can again be seen, 
e.g., in the framework of a schematic model. 
The second step (b) is RPA with
the coupling to the continuum, which  gives no width 
since the IAR is below the 
proton emission threshold. The aim of the third step (c) (inclusion of 
the spreading width) is 
to know whether the particle-phonon coupling model we 
have adopted can introduce some spurious width because of the approximations 
made: finite size of the set of 1p-1h plus phonon states, overcompleteness 
of the 1p-1h plus phonon basis, violation of the Pauli principle, simple 
form of the particle-phonon vertex function. The fact that 
we obtain only 4 keV for the spurious 
$\Gamma^\downarrow$ with the force SGII is very satisfactory since 
it means that the approximations
we have mentioned are safe in this case. 

In the case of the force SIII the spurious width (24 keV) is larger than  
that obtained with the force SGII. 
We have reached the conclusion that this spuriosity is due by half 
to the particle-phonon coupling and by half to the
fact that HF-TDA is not completely able to recover the isospin symmetry.  
To estimate the spuriosity due to the particle-phonon coupling, 
we have considered an ideal analog state at zero energy with the 
following schematic wave function: 
\begin{equation}
 |A\rangle = {1\over \sqrt{2T_0}}\ \hat T_- |HF\rangle 
	     = \sum_{\pi,\nu^{-1}} X_{\pi,\nu^{-1}} |\pi,\nu^{-1}\rangle,
\end{equation}
with $\nu^{-1}$ restricted to the excess neutrons and 
\begin{equation}
 X_{\pi,\nu^{-1}} = {1\over \sqrt{2T_0}}\ \delta(l_\pi,l_\nu)
 \delta(j_\pi,j_\nu)\ \sqrt{2j_\pi+1}\ \int dr\ u_\pi(r) u_\nu(r).
\end{equation}
We have calculated the coupling of this schematic IAR with 
the 1p-1h plus phonon 
states adopted for all the calculations of Table I, and we have obtained a 
state at 27.6 keV whose width is 13 keV. Therefore, 13 
keV is the broadening of the IAR introduced spuriously by the coupling with
particle-hole-phonon configurations. The remaining 11 keV 
is still larger than the value obtained with the interaction SGII. 
This is on one hand related to the different
single-particle levels obtained for the two interactions. As mentioned 
above, 
proton particle and neutron hole levels are more separated in the case 
of the force SIII. Consequently, the isospin breaking in 
the HF field is restored less efficiently by TDA and the energy of the 
IAR is less close to zero than in 
the case of the force SGII. Therefore, in the case of SIII we diagonalize  
the effective Hamiltonian (\ref{H_eff}) at a higher value of the energy $E$ and 
the spreading width of the IAR 
at higher excitation energy is larger. Moreover, 
the imaginary parts of the self-energy terms in our model 
would cancel exactly 
to give a zero spreading width if the single-particle radial wave 
functions of neutrons and protons with the same quantum numbers 
$(n,l,j)$ were identical. This cancellation can be seen by looking at the 
expressions of the four diagrams of 
Fig. 1~\cite{Col94}. However, the cancellation is not complete 
if the single-particle radial wave functions 
are different. The difference between 
radial wave functions is larger in the case of the force SIII, so again 
the spurious width is expected to
be larger than that of SGII. 

If we now include the Coulomb interaction between protons in the HF mean 
field, the proton levels are pushed up and become higher than the 
corresponding neutron levels. The energies of the unperturbed proton 
particle-neutron hole 0$^+$ configurations have positive values 
between 11 and 11.8 MeV for SGII and between 8.8 and 10.3 MeV 
for SIII (for the six main configurations already considered 
above). This difference between the two forces is essentially related 
to the fact that in the case 
of the force SGII the neutron
holes are more bound. 
Therefore, the proton particles which enter the IAR wave function
have less energy available ($\varepsilon_p = E_{IAR} + \varepsilon_h$) and a 
smaller probability to escape. This explains why the 
$\Gamma^\uparrow$ is considerably smaller than in the case of the force SIII. 

Finally, the most important result of our calculation is that the total width 
obtained by employing the force SIII (in the case of the 
complete calculation III, last column of Table 1) nicely agrees with the 
experimental finding $\Gamma_{TOT}^{(exp)}$ = 230 keV. The improvement with 
respect to calculation II (without CSB-CIB forces) is about 15\%. 
This shows that CSB-CIB forces can contribute significantly to 
the total width of the IAR. \\

\section{Comparison with the IVMR doorway state approach}
We have seen that 
microscopic approaches, like the particle-phonon coupling model 
described above or the 2p-2h TDA model of Ref.\cite{Ada_find} 
can give a reasonable description of IAR widths in spite 
of some sensitivity to the effective interactions. The spreading 
widths come from the coupling terms $Q_1HQ_2$ appearing in 
$W^{\downarrow}$ of Eq.\ (\ref{H_eff}), and this coupling between the 
simple $Q_1$ configurations and more complex $Q_2$ configurations is 
produced mostly by the isospin-conserving Skyrme interaction (in fact, this is 
the only residual interaction we keep in the calculation of $W^{\downarrow}$ 
in the previous sections). Thus, the fact that the resulting 
$\Gamma^{\downarrow}$ is non-zero is due entirely to the effects of 
the Coulomb and other isospin-breaking forces in the mean field as they 
produce a finite density of states with the isospin of 
the parent nucleus at the IAR energy. On 
the other hand, in the approaches of Refs.~\cite{Mek70,Aue83,Suz96} 
$\Gamma^{\downarrow}$ originates from the coupling of an ideal analog 
state $\vert A \rangle$ with a specific doorway state, namely the IVMR 
in the daughter nucleus, via the isospin-breaking part of the 
Hamiltonian (usually the isovector component of the Coulomb force). 
Here, we show that these apparently different points of view can be 
connected.

Let us first recall the expression for $\Gamma^{\downarrow}$ obtained 
in Ref.\cite{Suz96}. The 
Hamiltonian is assumed to be a sum of an isospin-conserving part plus 
the Coulomb interaction, $H = H_0 +V_C $, and the parent ground state 
which is eigenstate 
of $H_0$ is denoted by $\vert 0 \rangle$. The three isospin components 
of the IVMR in the daughter nucleus are schematically written as 
\begin{eqnarray}
\vert M; T_0 +i, T_0 -1 \rangle & = & \vert \{ \vert 0 \rangle^{T=T_0} 
\otimes \ |ph^{-1} \rangle^{T=1} \}^{T+i}_{T_0 - 1} \rangle~,
\label{eq1}
\end{eqnarray}
where $i=-1,0,1$ and $\vert ph^{-1} \rangle$ stands for a combination 
of monopole p-h excitations. In Ref.\cite{Suz96} the IAR spreading 
width $\tilde \Gamma^{\downarrow}$ (we use this notation for the 
spreading width calculated by following the doorway state 
approach) was expressed in terms of the 
analog state energy $E_A$, IVMR energies $E_M^{T_0 +i}$, the width 
$\Gamma_M (E_A)$ of the IVMR evaluated at energy $E=E_A$, and the 
reduced Coulomb matrix element:
\begin{equation}
 \tilde v_{C} = \frac{1}{\sqrt{3}}
 \langle (ph^{-1})^{T=1} \parallel V_C^{(1)} \parallel 0 \rangle~,
\label{not_suz}\end{equation}
where $V_C^{(1)}$ is the isovector part of the Coulomb potential. If 
one neglects the isospin splittings of the IVMR 
and adopts a common 
value $E_M^{T_0 +i} \simeq E_M$ the expression of $\tilde 
\Gamma^{\downarrow}$ takes the simple form
\begin{equation}
\tilde \Gamma^\downarrow_A = \Gamma_M(E_A) {\vert {\tilde v_C}\vert^2 \over
  (E_A - E_M)^2 + ({\Gamma_M\over 2})^2 }~.
\label{eq2}\end{equation}
Furthermore, it was shown that the isospin mixing probability of the 
$T_0 + 1$ component of IVMR in the parent ground state $|\pi\rangle$ 
is given in second order perturbation theory by
\begin{eqnarray}
 \vert c_{T_0+1}\vert ^2 & \equiv & {1\over 2(T_0+1)} \langle \pi |  T_-  T_+ 
 | \pi \rangle \nonumber \\
 & = & { 1 \over T_0+1 }{ \vert{\tilde v_C}\vert^2 \over \vert \Delta 
E_M \vert ^2 }~,
\label{eq3}
\end{eqnarray}
where $\Delta E_M$ is the excitation energy of the IVMR in the parent 
nucleus which is approximately equal to $E_A - E_M$ of 
Eq.\ (\ref{eq2}). We can safely neglect $\Gamma_M$ in Eq.\ (\ref{eq2}) 
and thus obtain
\begin{equation}
\tilde \Gamma^\downarrow_A  
 \sim {1\over 2} \Gamma_M(E_A) \langle \pi |
  T_-  T_+ | \pi \rangle~.
\label{gamma1}\end{equation}

In the microscopic models~\cite{Ada_find,Col94} 
the spreading width of the IAR results from the
couplings mediated by the 
isospin-conserving operator $W^\downarrow$ defined in section II. 
Denoting the RPA eigenstate corresponding to the 
IAR by $|A\rangle$, we can write its width as 
\begin{equation}
 \Gamma^\downarrow_A  = -2\ Im\ \langle A | W^\downarrow | A
 \rangle.
\label{gamma2_1}\end{equation}
The IAR wave function can be well approximated 
by 
\begin{equation}
 |A\rangle = {1\over \sqrt{2T_0}}\  T_- | \pi \rangle~. 
\label{schem_wf}\end{equation}
For the isospin-conserving $W^\downarrow$
interaction we make the ansatz: 
\begin{eqnarray}
 W^\downarrow & = & a({{\vec T}}\cdot{{\vec T}} - b) \nonumber \\
 & = & (a_R + i a_I)({{\vec T}}\cdot{{\vec T}} - b)~,
\label{schem_Wdown}\end{eqnarray}
and obtain
\begin{eqnarray}
 \langle A | W^\downarrow | A \rangle & = & {a\over {2T_0}} \langle
  \pi |  T_+ ({ {\vec T}}\cdot{ {\vec T}} - b)  T_- 
 |  \pi \rangle
 \hfill\nonumber \\
 \ & = &  {a\over {2T_0}} \langle  \pi |  T_+  T_- ({ {\vec
 T}}\cdot{ {\vec T}} - b) 
 |  \pi \rangle
 \hfill\nonumber \\
 \ & \approx & 
 {a\over {2T_0}} \langle  \pi |  T_+  T_- |  \pi
 \rangle \langle  \pi\vert ({ {\vec T}}\cdot{ {\vec T}} 
 - b) | \pi \rangle~.
\label{15_of_fax}\end{eqnarray}
In the last step of the above equation the contributions of the 
excited states to the closure relation have been dropped because they 
are of order $\vert c_{T_0 +1}\vert ^2$. Let us introduce $\tilde T$ 
by: 
\begin{equation}
 \langle  \pi | { {\vec T}}\cdot{ {\vec T}} 
 |  \pi \rangle \equiv
 \tilde T (\tilde T + 1)~,
\label{ttilde}\end{equation}
($\tilde T$ differs slightly from $T_0$ because $\vert \pi \rangle$ has 
isospin mixing) and choose $b={\tilde T}^2$. Then:
\begin{equation}
 \langle A | W^\downarrow | A \rangle = a [ \tilde T (\tilde T + 1) - b ]
 = a\ \tilde T. 
\label{end2}\end{equation}
Thus, the two expressions (\ref{gamma1}) and (\ref{gamma2_1}) are 
equal if the following condition is satisfied:
\begin{equation}
  a_I = -{\Gamma_M(E_A)\over 4\tilde T} \langle \pi |  T_-  T_+ | 
 \pi \rangle.
\label{end3}\end{equation}

Next, we diagonalize the TDA schematic model with the interaction
\begin{equation}
 v_{eff} = (a_R + i a_I) ({ {\vec T}}\cdot{ {\vec T}} - b)~. 
\label{veff}\end{equation}
The complex eigenvalue is
\begin{eqnarray}
 E_A - i {\Gamma_A \over 2} & = & \varepsilon_{ph} + {v_{eff}\over 2} 
\sum_{ph} | \langle 
 (ph^{-1})^{T=1,T_z=-1} |  T_- | 0 \rangle |^2 \hfill\nonumber \\
 \ & = & \varepsilon_{ph} + {a_R\over 2}\cdot 2T_0 + i {a_I\over 2}\cdot 
2T_0~,
\label{schem_TDA}\end{eqnarray}
where $\varepsilon_{ph}$ is the degenerate unperturbed energy
of the 0$^+$ proton particle-neutron hole configurations. Replacing 
$a_I$ by its value (\ref{end3}) we see that $\Gamma_A$ becomes  
\begin{equation}
 \Gamma_A  = {1 \over 2}{\Gamma_M(E_A)}{T_0 \over \tilde T}
  \langle \pi | \hat T_- \hat T_+ | \pi \rangle~,
\label{fine3}\end{equation}
which is consistent with Eq. (\ref{gamma1}). 
Thus, the adopted interaction 
$W^\downarrow$ results in a value of the spreading width coming from 
the coupling of the IAR 
with states of 2p-2h type through the nuclear interaction,  
which is comparable
with the result of the IVMR doorway state approach in which the spreading  
width is obtained through the coupling of the IAR 
to the IVMR due to the Coulomb 
interaction. 
It is left as a future problem
the justification of the choice of (\ref{schem_Wdown}) from a microscopic 
study of optical potentials.    

\section{Conclusion}
Within the framework of a microscopic model based on self-consistent 
HF-RPA plus coupling with continuum configurations as well as with 
1p-1h plus phonon configurations, 
we have calculated the total width of the IAR in 
$^{208}$Bi. We have shown that if the nuclear isospin-breaking forces 
of CSB and CIB type are included in the Hamiltonian in addition to the 
Coulomb interaction, the width of the IAR is increased by 15-20\%. 
Thus, the nuclear isospin-breaking interactions which were already 
known to increase the isospin mixing of ground states have also 
significant contributions to the total width of the IAR. As far as 
comparison with experiment is concerned, the values of 
$\Gamma^{\downarrow}$ and $\Gamma_{TOT}$ calculated with SIII are in 
satisfactory agreement with the data whereas those obtained with SGII 
are not so good due to the peculiarities of the single-particle spectra 
of SGII. 

Our microscopic model introduces some spuriosity in the evaluation 
of the total width of the IAR. This spuriosity turns out to be quite small 
and we have shown that it is due partly 
to the incomplete restoration of symmetry by TDA 
and partly to the 1p-1h plus phonon model. Indeed, this model does not 
have the full self-consistency of a second RPA calculation which would 
be free in principle of spurious isospin violations. However, a second 
RPA calculation including also continuum effects would be extremely 
difficult and it has never been done so far. 

Finally, we have been able to propose for the first time a connection 
between the microscopic   
model and IVMR doorway state approaches for 
the spreading width of the IAR. 
This connection is possible 
by making a phenomenological 
ansatz for the isospin dependence of the nucleon optical potential. 

\section*{Acknowledgments}
G.C. likes to acknowledge the nice hospitality of the Division de Physique 
Th\'eorique (IPN, Orsay) where part of the work has been done. H.S. and 
P.F.B. acknowledge the warm hospitality of the Institute for Nuclear Theory 
of Seattle, where this project started during the 1995 workshop on nuclear
structure.

\newpage

\begin{figure}
\caption{ Diagrams corresponding to the coupling of the 1p-1h  
configurations to the more complicated states including a phonon (wavy line). 
The sum of the four diagrams gives the matrix element 
W$^\downarrow_{ph,p^\prime h^\prime}$ 
quoted in the text. }
\end{figure}

\widetext

\begin{table}
\caption{IAR results with 
the interaction SIII. Three different types of interactions based
on SIII are used 
in the calculations: I) Skyrme interaction
without Coulomb force; II) Skyrme interaction with Coulomb
force; III) Skyrme interaction with Coulomb
force and CSB-CIB forces.  Three different microscopic models are
also adopted: a) TDA without the coupling to the continuum; b) RPA
with the coupling to the continuum; c) RPA with the couplings to
both the continuum and the phonons.  Energies are given in MeV and
widths 
are in keV. The percentage of total 
strength m$_0$=(N-Z)/2 exhausted by the IAR is also shown.} 
\vspace{0.3cm}
\begin{tabular}{cccccccccc}
   & \multicolumn{2}{c}{(a) Discrete TDA} & \multicolumn{3}{c}{(b) 
   RPA +    W$^\uparrow$} & 
   \multicolumn{4}{c}{(c) RPA + W$^\uparrow$ + W$^\downarrow$} \\
\tableline 
   & E$_{IAR}$ & \% of m$_0$ & E$_{IAR}$ & $\Gamma^\uparrow$ & \% of m$_0$ &
   E$_{IAR}$ & $\Gamma_{TOT}$ & $\Gamma^\downarrow$
   & \% of m$_0$ \\ 
\tableline
 I   & 0.268 & 99.9 & -     & -   & -  & 0.267 & 24  & 24  & 99.7
 \\
 II   & 18.50 & 85   & 18.50 & 124 & 97 & 18.36 & 194 & 70  & 97 
 \\
                   & 18.28 & 16   & & & & & & & 
 \\
 III  & 18.64 & 80   & 18.65 & 128 & 96 & 18.54 & 228 & 100 & 96 
 \\
                   & 18.39 & 11  & & & & & & & \\
\end{tabular}
\end{table}     
  
\vspace{2.0cm}

\newpage

\begin{table}
\caption{IAR results obtained with the interaction
SGII. For details, see the caption to the previous table.} 
\vspace{0.3cm}
\begin{tabular}{cccccccccc} 
   & \multicolumn{2}{c}{(a) Discrete TDA} & \multicolumn{3}{c}{(b) 
   RPA +    W$^\uparrow$} &
   \multicolumn{4}{c}{(c)  RPA + W$^\uparrow$ + W$^\downarrow$} \\
\tableline
   & E$_{IAR}$ & \% of m$_0$ & E$_{IAR}$ & $\Gamma^\uparrow$ & \% of m$_0$ &
   E$_{IAR}$ & $\Gamma_{TOT}$ & $\Gamma^\downarrow$
   & \% of m$_0$ \\ 
\tableline
 I    & 0.185 & 99.8 & -   & -  & -  & 0.185 & 4   & 4  & 99.8  
 \\
 II   & 18.50 & 87 & 18.61 & 40 & 96 & 18.52 & 138 & 98 & 95 
 \\
 III  & 18.65 & 87 & 18.77 & 42 & 96 & 18.69 & 164 & 112 & 96 \\
\end{tabular}
\end{table}
  
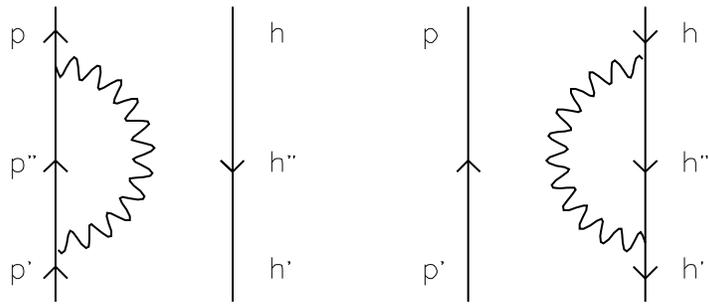
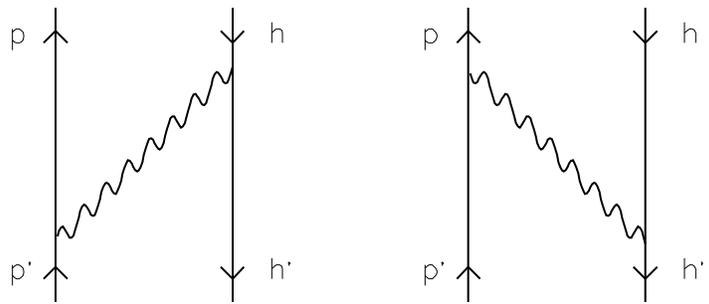
\end{document}